\documentclass[aps,pre,showpacs,twocolumn]{revtex4-1}

\usepackage{graphicx}
\usepackage{color}
\bibstyle{apsrev.bib}

\newcommand{\mean}[1]{{\left< #1 \right>}}
\newcommand{\bino}[2]{{#1 \choose #2}}

\begin{document}

\title{Interacting Elastic Lattice Polymers: a Study of the Free-Energy of Globular Rings}

\author{M. Baiesi and  E. Orlandini}
\address{Department of Physics and Astronomy, University of Padua, Via Marzolo 8, I-35131 Padova, Italy}
\address{INFN, Sezione di Padova, Via Marzolo 8, I-35131 Padova, Italy}

\begin{abstract}
We introduce and implement a Monte Carlo scheme to study the equilibrium 
statistics of polymers in the globular phase. It is based on  
a model of ``interacting elastic lattice polymers'' and allows a sufficiently good 
sampling of long and compact configurations, an essential prerequisite to 
study the scaling behavior of free energies.
By simulating interacting self-avoiding rings  at several temperatures in the collapsed phase,
we estimate both the bulk and the surface free energy. 
Moreover from the corresponding  estimate of the entropic exponent $\alpha-2$ we provide evidence that, 
unlike for swollen and $\Theta$-point rings, 
the hyperscaling relation is not satisfied for globular rings.
\end{abstract}

\pacs{
64.60.De   
36.20.-r, 
05.10.Ln,  
64.60.an,  
}

\maketitle

\section{Introduction}

It is  well known that diluted polymers in bad solvent conditions 
may undergo a conformational transition from the swollen (extended) phase to a globular 
one~\cite{DeGennes:1979,Vanderzande:1998,JansevanRensburg_book}. 
In the last decades this so-called \emph{collapse transition} has been the 
subject of numerous experimental and theoretical studies that have been essentially 
focused either on the swollen phase or on the nature and scaling properties 
of the collapse ($\Theta$) transition~\cite{DuplantierSaleur_PRL87:Theta2d,Seno_Stella:1988,GrassbergerHegger_JP95:theta2d,Tesi_et_al:1996:J_Stat_Phys,Iwata:1989:Macromol}. 
On the other hand very little is known about polymers
in the globular 
phase~\cite{Owczarek_PB_PRL93:collapsed,Duplantier_PRL93:comment,Owczarek_PB_PRL93:reply,Brak:1993:JPA,Grassberger:PRE:2002,Baiesi:2006:PRL,Baiesi:2007:PRL}, 
despite the study of their statistics can help in understanding 
the conformational and thermodynamic properties of more complex forms of compact 
polymers such as proteins in native states.

If one is mainly interested in the scaling behavior of either metric or thermodynamics 
observables (including critical exponents) of collapsing polymers, a good model to 
consider is the class of $N$-site interacting self-avoiding walks (ISAWs), namely, 
non self-intersecting walks of $N$ sites on Bravais lattices with an 
attractive interaction between adjacent non-bonded sites. Indeed, by tuning this attraction 
in terms of an effective temperature $T = 1/\beta$, ISAWs display a collapse  transition 
towards a globular phase whose thermodynamics is governed, in the limit of large $N$, 
by the scaling behavior of the partition 
function~\cite{Owczarek_PB_PRL93:collapsed,Duplantier_PRL93:comment,Owczarek_PB_PRL93:reply,Brak:1993:JPA,Grassberger:PRE:2002,Baiesi:2006:PRL,Baiesi:2007:PRL}
\begin{equation}
Z_N(\beta) \sim \mu^N e^{-\sigma N^{(d-1)/d}} N^\theta \left ( 1+ B N^{-\Delta}\cdots \right ).
\label{ZN}
\end{equation}
(such structure is also supported by theoretical results for partially directed 
ISAWs~\cite{Brak:1992:JPA,Nguyen:2013:JSP} and for 
dense polymers~\cite{DuplantierSaleur_NPB87:dense,DuplantierDavid_JSP88}).
In~(\ref{ZN}) the connective constant $\mu$ and the amplitude $B$ of the correction to scaling 
are model dependent quantities  and function of  $\beta$,
whereas the (entropic) exponent $\theta$ is a universal (i.e.~model independent) critical 
exponent whose value depends only on dimension ($d$)
and on polymer topology (for instance linear, circular, knotted etc).
Note in~(\ref{ZN}) the presence of the term $e^{-\sigma N^{(d-1)/d}}$ that
describes the surface penalty contribution ($\sigma>0$ for $\beta>\beta_\Theta$) related to the
higher free energy acquired by the monomers exposed to the 
solvent~\cite{Owczarek_PB_PRL93:collapsed,Brak:1993:JPA,Grassberger:PRE:2002,Baiesi:2006:PRL,Baiesi:2007:PRL}.
The presence of this surface renders the asymptotic analysis of the thermodynamics of 
globular polymers  more complicated than that for dense polymers,  
at least for two reasons: First, the presence of the additional  
unknown parameter $\sigma$ increases the 
complexity of the analysis of the scaling law~(\ref{ZN}). Second,
there is  strong numerical evidence that the surface generates strong corrections 
to scaling when linear polymers are considered~\cite{Baiesi:2006:PRL}. The reason is that configurations 
with one or both ends on the surface of the globule surface have different entropic 
exponents with respect to  those where both ends are in the interior
of the globule, the latter become asymptotically relevant only for long chains.
This effect has been observed for $d=2$ globules~\cite{Baiesi:2006:PRL} and 
should be even more pronounced for globular polymers in $d=3$,
where the surface/volume ratio is larger than in two dimensions.  A way to
get around this problem  consists for instance in  looking at globular rings.

Scaling laws such as~(\ref{ZN}) are usually studied numerically by $N$-varying (grand-canonical)  Monte Carlo 
approaches~\cite{Berg&Foester:1981:Phys-Lett-B,Aragao&Caracciolo:1983:J-Physique,
Grassberger:1997:PRE,Rechnitzer:2008:JPA}.  
An example is the so called 
BFACF algorithm~\cite{Berg&Foester:1981:Phys-Lett-B,Aragao&Caracciolo:1983:J-Physique}
(from the initials of the authors), 
a set of local moves including the deletion/addition of a 
few monomers along the walk.
Being ergodic within the class of rings with a
given knot type~\cite{Janse-van-Rensburg&Whittington:1991b:J-Phys-A}, this algorithm
has also been  extensively used  to study the effect of topological constraints on the 
asymptotic properties of knotted rings in the swollen phase~\cite{Orlandini:1998:J-Phys-A}.
However, it is known that its straightforward application
to the globular phase fails to reproduce the $N$-varying statistics with 
a controlled average number of monomers $\mean{N}$ if  $N$ is sufficiently 
large~\cite{Marcone:2007:PRE,Baiesi_et_al:2011:PRL}, 
a condition to achieve for studying the scaling law~(\ref{ZN}) properly.

Here we introduce a stable numerical scheme to study the large $N$ behavior of globular polymers. 
This is applied on a model of interacting polymer
that from now on we refer to as an \emph{interacting elastic lattice polymer} (IELP). 
The IELP can be seen as an extension of the elastic lattice polymer (ELP), a model 
that has been introduced to study dynamical properties 
of linear polymers~\cite{Heukelum:2003:JChemPhys,Wolterink:2006:PRL}
and more recently extended to investigate the equilibrium  properties of rings 
in the swollen phase~\cite{Baiesi_et_al:2010:PRE}. 
Briefly (see the next section for a better definition of the model) an 
$M$-monomer IELP is an ISAW where locally the self-avoiding 
condition can be partially relaxed by allowing some consecutive monomers to condense on the
same lattice site.  In this respect the $M$ monomers of an IELP are elastically stored along an 
ISAW backbone with fluctuating length $N\le M$. 
This additional degree of freedom, not present in the standard 
ISAW model, can then be tuned to better stabilize 
the average length $\mean{N}$ when $N$-varying algorithms are used  in stochastic sampling.
Moreover, by restricting the sampling to circular IELPs (rings), the aforementioned 
finite-size corrections, due to the positions of the  polymer ends with respect to the globule 
surface, can be avoided.  Finally, by 
looking at the asymptotic behavior  of the equilibrium averaged stored length density 
\begin{equation}
\rho_M = \frac{M-\mean{N}}{M},
\label{rhoM}
\end{equation}
we  have access to the full scaling law~(\ref{ZN}) and we are able to 
estimate its parameters in the globular phase.

This paper is organized as follows: In Section~II we introduce the IELP model 
and its asymptotic behavior  in the globular phase.
In particular we show how to extract information on 
the $\mu$, $\sigma$, and $\theta$ exponents by extrapolating 
the average stored length density $\rho_M$ for large $M$ values. 
In Section~III we describe the $N$-varying  Monte Carlo algorithm used to sample closed IELPs 
on the hypercubic lattice and its implementation on a  Multiple Markov Chains scheme, 
an additional stochastic procedure known to increase the mobility of the sampling 
for compact configurations.  In Section~IV we show the results obtained for collapsed
rings, ending with some conclusions in Section~V.

\section{The IELP Model for globular polymers}
A commonly used model to study numerically the polymer collapse transition  is the ISAW, namely a
SAW augmented with an attractive energy $-E$ equal to the number of contacts 
(non-bonded pairs of nearest-neighbors sites of the SAW).

In the grand-canonical ensemble the 
large $N$ behavior of ISAWs is governed by the singularities of 
the generating function
\begin{equation}
G(K,\beta) = \sum_N K^N Z_N(\beta)
\label{gen_1}
\end{equation}
where  $K$ is the \emph{step fugacity}, $\beta$ the inverse of the temperature
and
\begin{equation}
Z_N(\beta) = \sum_{E} e^{-\beta E} C_N(E)
\end{equation}
is the canonical partition function associated with the number of $N$-sites ISAWs with energy $E$, $C_N(E)$.  
The singularities of~(\ref{gen_1}) develop as $K\to K_c(\beta)^{-}$ and 
Monte Carlo algorithms based on the BFACF moves 
must bring the value of $K$ as close as possible to 
$K_c(\beta) = 1/ \mu(\beta)$ to get a good statistics of  configurations with large $N$.

Unfortunately a straightforward application of the BFACF algorithm in the globular phase
meets a difficulty.  Indeed, due to the presence of the surface
term $\sim e^{-\sigma N^{2/3}}$ in~(\ref{ZN}) of $Z_N$, the grand-canonical average
$ \langle N \rangle = \Sigma_N N K^N Z_N / \Sigma_N K^N Z_N$ does not grow continuously
to $+ \infty$ as $K\to K_c(\beta)^{-}$ but instead jumps discontinuously to infinity 
right at $K=K_c$, making the sampling of long walks quite unrealistic~\cite{Marcone:2007:PRE}.

\begin{figure}[!bt] 
\begin{center}
\includegraphics[width=8cm]{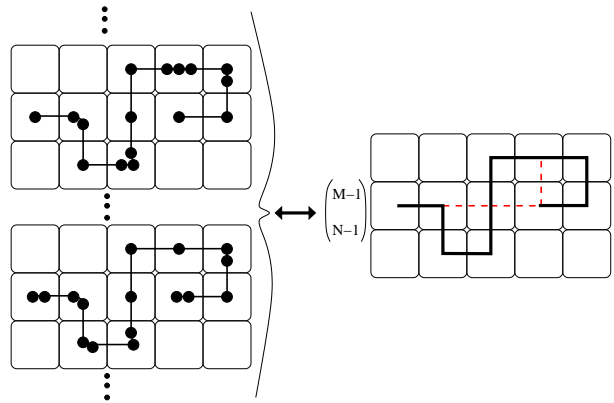}
\end{center}
\caption{Example of ELPs with $M=16$ monomers and $N=10$ sites (left) sharing
the $N$-sites SAW silhouette (right). Dashed lines represent the 
energetic contacts associated to that configuration.}
\label{fig01} 
\end{figure}

In order to allow the number of sites $N$ to fluctuate and grow in a rather controlled way
we introduce the IELP model, an extension of 
the ELP model for SAWs in which self-attracting interaction is taken into account.
As in the ELP model, the idea is to accommodate $M \ge N$ monomers on 
an underlying $N$-site SAW silhouette
drawn on a lattice. To maintain the connectivity constraint of a polymer chain,
at least one monomer resides in each site of the underlying  $N$-site SAW and 
only consecutive monomers can share the same lattice site (see Fig.~\ref{fig01}).

Certainly one could sample IELPs by updating the position of all $M$ monomers 
within the lattice~\cite{Schram:JCP:2013}.
However, this level of detail is not necessary when the energy function depends only on the 
$N$-sites ISAW silhouette: in this case it is sufficient
to sample directly the ISAWs backbone with rates that take into
account the degeneracy  of each ISAW configuration.
Since for each $N$-site ISAW there are  $\bino{M-1}{N-1}$
possible IELPs,
the total weight associated with an IELP configuration with $M$ monomers, $N$ lattice sites 
and $E$ contacts, is given by
\begin{equation}
W =K^{N} e^{-\beta E}\bino{M-1}{N-1}.
\label{weight}
\end{equation}
By summing over all possible values of $E$ and $N$ compatible with $M$ 
we obtain the generating function of IELPs with $M$ monomers  
\begin{equation}
Z^{IELP}(M,\beta) = \sum_{N\le M} \bino{M-1}{N-1} K^N Z_N(\beta). 
\label{Z_IELP}
\end{equation}

Note that, when $M$ and $K$ are both kept fixed, the number of sites $N$ oscillates 
around an average $\mean{N}_{K,M,\beta}$ that scales linearly with $M$, 
with fluctuations of the order of $\sqrt{M}$.  
This property, as we will show below, is very helpful in sampling 
long walks undergoing a $\Theta$-collapse. 

A second advantage in sampling IELPs is the possibility of estimating $\mu(\beta)$, $\sigma(\beta)$ 
and $\theta$ by computing the asymptotic of the  mean stored length 
density~(\ref{rhoM}) within a saddle-point approximation~\cite{Baiesi_et_al:2010:PRE}.  
To show this we first notice that, by defining
$\omega(\beta) \equiv \mu(\beta) K$ 
and substituting the scaling~(\ref{ZN})
in~(\ref{Z_IELP}), the average $\mean{N}$ can be
computed by using the relation
\begin{equation}
\mean{N}(\beta) = \omega \frac{\partial}{\partial \omega} \log Z^{IELP}(M,\beta).
\end{equation}
If the partition function of ISAWs were simply $Z_N (\beta)=\mu(\beta)^N$, one would get the 
exact expression
$N_\infty(\beta) = \mean{N}(\beta) = M \omega(\beta)/[1+\omega(\beta)]$~\cite{Baiesi_et_al:2010:PRE}.
Thus, by using the  change of variable  $N = x N_{\infty}$, the saddle-point approximation
of the partition function  becomes 
\begin{equation}
Z^{IELP}(M,\beta) \sim (1+\omega)^M \sqrt{\frac{\omega M}{2\pi}} 
\int_{-\infty}^{+\infty} e^{M \omega F(x)} d x
\end{equation}
with 
\begin{eqnarray}
F(x) &=& -\frac{(x-1)^2}{2} + \frac{1}{M\omega}\left[ -\sigma (x N_{\infty})^{2/3} \right. \\
 && \ \ \left.  + \theta\ln (x N_{\infty}) + \ln \left (1+ B (x N_{\infty})^{-\Delta} \right)\right] \nonumber
\end{eqnarray}
Denoting by $\bar{x}$ the location of the maximum of $F(x)$, the approximation gives
\begin{equation}
Z^{IELP}(M,\beta) \sim (1+\omega)^M \frac{e^{M\omega F(\bar{x})}}{\sqrt{|F''(\bar{x})|}}
\end{equation}
Since in the large $M$ limit 
\begin{eqnarray}
|F''(\bar{x})| &=& 1 + {\cal{O}}\left (\frac{1}{M}\right),\\
\bar{x} &=& 1 + {\cal{O}}\left (\frac{1}{M}\right),
\end{eqnarray}
we get 
\begin{equation}
M \omega F(\bar{x}) = -\sigma N_{\infty}^{2/3} + \theta \ln N_{\infty} + \ln \left (1+ B (N_{\infty})^{-\Delta} \right ).
\end{equation}
We then have
\begin{widetext}
\begin{eqnarray}
\langle N \rangle  =  \omega \frac{\partial}{\partial \omega} \ln Z^{IELP}(M,\beta)
&=& N_{\infty} \left [1 - \frac{2\sigma}{3} \frac{N_{\infty}^{-1/3}}{1+\omega} + \frac{\theta}{N_{\infty}} \frac{1}{1+\omega}- \frac{B \Delta N_{\infty}^{-\Delta-1}}{1+B N_{\infty}^{-\Delta}} \frac{1}{1+\omega} \right ] \nonumber \\
&=& N_{\infty} \left [1 - \frac{2\sigma}{3} \frac{N_{\infty}^{-1/3}}{1+\omega} 
+ \frac{\theta}{M\omega} - \frac{B \Delta }{M\omega} \left ( \frac{1+\omega}{M\omega}\right )^{\Delta} \left ( 1+BN_{\infty}^{-\Delta} \right )^{-1} \right ] 
\end{eqnarray}
from which we get the scaling of the density $\rho_M(\beta)=1-\mean{N}/M$ to order $1/M^{1+\Delta}$,
\begin{equation}
\rho_M(\beta)= \rho_{\infty}(\beta) 
\left[ 
1 + 
\frac{2\sigma(\beta)}{3} \left(\frac{\omega(\beta)}{1+\omega(\beta)}\right)^{2/3} \frac{1}{M^{1/3}} 
- \frac{\theta}{M} 
+  B(\beta) \Delta \left( \frac{\omega(\beta)}{1+\omega(\beta)}\right)^{-\Delta}  \frac{1}{M^{1+\Delta}} \right]
\label{rho-fit}
\end{equation}
\end{widetext}
where 
\begin{equation}
\rho_{\infty}(\beta) = 1-\frac{N_\infty}{M} = \frac{1}{1+\omega(\beta)}.
\end{equation}

We stress that at high temperature the surface term is not present and the 
leading term is the one proportional to the exponent $\theta$, which scales as $1/M$~\cite{Baiesi_et_al:2010:PRE}.
On the other hand, in the globular phase the $\sigma$ term is $\sim 1/M^{1/3}$: 
in the limit of large $M$ it dominates the entropic one, 
which now acts as  a (weak) correction to scaling. 

\section{Monte Carlo algorithm for IELP}
Since the Monte Carlo algorithm for IELPs is based on the BFACF moves, its 
ergodicity is a direct consequence of the one already proven for BFACF~\cite{Madras&Slade:1993}.
The Monte Carlo scheme  samples ISAWs configurations according to the 
statistical weight~(\ref{weight}) of IELPs;  it is then enough to require that
the rate of jump $R_{AB}$ from a configuration $A$
to a configuration $B$ satisfies the detailed balance condition
$W_A R_{AB} = W_B R_{BA}$. From  Eq.~(\ref{weight}) we get
\begin{eqnarray}
\frac{R_{AB}}{R_{BA}} &=& \frac{W_B}{W_A} \\
&=&K^{N_B-N_A}  e^{-\beta (E_B - E_A)}\bino{M-1}{N_B-1} / \bino{M-1}{N_A-1} \nonumber \\
&=&K^{N_B-N_A}  e^{-\beta (E_B - E_A)}\frac{(N_A-1)!(M-N_A)!}{(N_B-1)!(M-N_B)!} \nonumber \,.
\label{db}
\end{eqnarray}

\begin{figure}[!b] 
\begin{center}
\includegraphics[width=5cm]{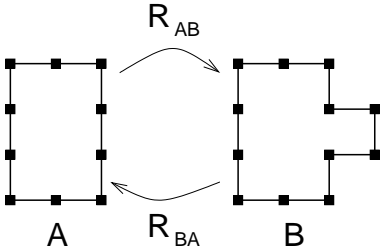}
\end{center}
\caption{Example of a proposed move (and its inverse) that transforms locally 
an $N$-site configuration $A$ to an $(N+2)$-site configuration  $B$.}
\label{fig02} 
\end{figure}

We now specialize the discussion to $M$-monomers circular IELPs, whose underlying silhouette 
is given by a $N$-steps self-avoiding polygon (SAP) on a $d$-dimensional 
hyper-cubic lattice. Note that in this case the periodicity in monomer labeling 
requires $N+1 \equiv 1$ with the number of monomers equal to the number of steps. 
The BFACF moves that need some care in their implementation are those 
changing the number $N$ of monomers.
These are based on the addition/removal of a crankshaft bulge: in such a motif 
a site $i$ along the chain is a nearest neighbor of site $i+3$ and, together
with $i+1$ and $i+2$, they form a unit square on the hypercubic lattice 
(see a $d=2$ example in  Fig.~\ref{fig02}).

The addition/removal of two sites connects the ensemble of SAPs with $N_A=N$ sites to that
of SAPs with $N_B=N_A+2$ sites. Accordingly, equation~(\ref{db}) simplifies to
\begin{eqnarray}
\frac{R_{AB}}{R_{BA}} &=&K^{2} e^{-\beta (E_B - E_A)} \frac{(N-1)!(M-N)!}{(N+1)!(M-N-2)!} \nonumber \\ 
&=&K^2 e^{-\beta (E_B - E_A)}\frac{(M-N)(M-N-1)}{(N+1)N}
\label{db1}
\end{eqnarray}
In practice a local update $A\to B$ that proposes the addition of two monomers 
is implemented by choosing, with uniform probability $P_A^{\rm ind} = 1/N$, 
one of the $N$ bonds of polygon A. Then one of the $2d-2$ possible directions orthogonal 
to the chosen bond, $(i,i+1)$,  is picked up with probability $P_A^{\rm dir} = 1/(2d-2)$.

The reverse update $B\to A$ is implemented accordingly
by choosing a bond of the polygon B with probability $P_B^{\rm ind} = 1/(N+2)$. 
By calling ${\cal A}_{AB}$ and ${\cal A}_{BA}$ the acceptance probability 
of the direct ($A\to B$) and reverse ($B\to A$) moves respectively,
the jump rates  $R_{AB}$ and $R_{BA}$ can be written as
\begin{eqnarray}
R_{AB} &=& f_{AB} P_A^{\rm ind} P_A^{\rm dir} {\cal A}_{AB} =\frac{f_{AB}}{2d-2} \frac{1}{N} {\cal A}_{AB} \nonumber \\ 
R_{BA} &=& f_{BA} P_B^{\rm ind} {\cal A}_{BA} = f_{BA} \frac{1}{N+2} {\cal A}_{BA}
\end{eqnarray}
where $f_{AB}$ and $f_{BA}$ are the frequencies with which the direct and reverse moves 
are proposed, respectively. These frequencies can be chosen to satisfy the detailed 
balance~(\ref{db}).
By choosing $f_{AB}=2d-2$ and $f_{BA}=1$  we get
\begin{eqnarray}
R_{AB} &=&\frac{1}{N} {\cal A}_{AB} \nonumber \\ 
R_{BA} &=&\frac{1}{N+2} {\cal A}_{BA}.
\end{eqnarray}
With this choice the acceptance probabilities become
\begin{eqnarray}
\frac{{\cal A}_{AB}}{{\cal A}_{BA}} &=& 
 \frac{N}{N+2}\frac{R_{AB}}{R_{BA}}\nonumber \\
&=&\frac{N}{N+2} K^2  e^{-\beta (E_B-E_A)} \frac{(M-N)(M-N-1)}{(N+1)N} 
\nonumber \\ 
&\equiv& K^2 e^{-\beta (E_B - E_A)}q(M,N)
\label{acc1}
\end{eqnarray}
where we have defined the ratio
\begin{eqnarray}
q(M,N)=\frac{(M-N)(M-N-1)}{(N+1)(N+2)}.
\end{eqnarray}
Note that $q(M,N)\le 1$ for $N\ge (M-2)/2$ and  $q(M,N)> 1$  otherwise.

In the spirit of a Metropolis criterion, the idea is to maximize both ${\cal A}_{AB}$ 
and ${\cal A}_{BA}$ compatibly with the above constraint on $q(M,N)$. 
Ideally, one of the two should be equal to $1$.
The following acceptance rates of the proposed move $N\to N+2$ and $N\to N-2$
satisfy the above requirements for any value of $N$:
\begin{eqnarray}
{\cal A}(E_A,N\to E_B,N+2)  = &&\  \min\{1,K^{2}\}  \times   \\
&&\ \min\{1,q(M,N)\}\times \nonumber  \\
&&\ \min\{1, e^{-\beta (E_B-E_A)}\} \nonumber \\
\label{acc3a}
{\cal A}(E_B,N\to E_A,N-2)  = &&\ \min\{1,K^{-2}\} \times  \\
 &&\ \min\{1, q^{-1}(M,N-2)\}  \times \nonumber  \\
 &&\  \min\{1,e^{-\beta (E_A - E_B)}\}. \nonumber 
\label{acc3b}
\end{eqnarray}
Hence, for a $d$-dimensional hyper-cubic lattice, 
when the frequency of proposed moves that increase the number of steps 
from $N$ to $N+2$, $f_{AB}$, is $2d-2$ times the frequency
of the reverse moves ($f_{BA}$), we can apply the filters (23)-(24) 
to accept them.
For an efficient implementation of the algorithm once, say,  the move $N\to N+2$
is proposed, it is more convenient to pass first the filter equal to 
$\min\{1,K^{2}\} \min\{1,q(M,N)\}$ and then, after the self-avoidance test is passed,
accept the move with probability equal to $\min\{1, e^{-\beta (E_B-E_A)}\}$.

In addition to the $N$-varying moves the Monte Carlo scheme provides
local corner flips~\cite{Madras&Slade:1993} and 
two-point pivot moves~\cite{Madras&Slade:1993}, 
the latter being essential to sample rings with arbitrary topology.

The algorithm just described designs a Markov Chain on the space of IELPs with fixed 
parameters $\beta$ and $M$. Several runs at different values of $M$ are then necessary to
span a broad range of density values at fixed $\beta$. An efficient 
way  to get several simulations with different $M$ at once 
can be obtained by embedding the above algorithm in a sampling scheme where 
several Markov chains at different $M$s  run in parallel, and  swaps 
of configurations between contiguous (in $M$)  Markov chains are performed stochastically.
This super Markov chain is known as \emph{Multiple Markov Chains} and in the past has been
proved to be very effective in increasing the sampling 
efficiency~\cite{Meyer:1991,Tesi_et_al:1996:J_Stat_Phys,Orlandini:1998:IMA}.

By considering (\ref{weight}) one can determine the acceptance ratio of configurations
swapping between two  Markov chains respectively of parameters $M$ and $M'$: 
since $\beta$ and $K$ are fixed, the only relevant term in (\ref{weight}) is the factorial and
the acceptance probability of the swap move  
$A =\{ (M,N) \& (M',N') \} \to B =\{ (M,N') \& (M',N) \}$ reduces to
\begin{equation}
{\cal A}_{AB} = \min\left\{1,\, \frac{(M-N)!(M'-N')!}{(M-N')!(M'-N)!}\right\}.
\end{equation}
If $M'>M$ this equation can be written as
\begin{equation}
{\cal A}_{AB} = \min\left\{1, \prod_{i=0}^{M'-M-1} \frac{M'-N'-i}{M'-N-i}\right\}
\end{equation}
Since each term of the product is larger than $1$ for $N>N'$, 
a swap that move the longer configuration to  higher values of $M$ is always accepted.

\begin{figure}[!tb]
\begin{center}
\includegraphics[width=7.8cm]{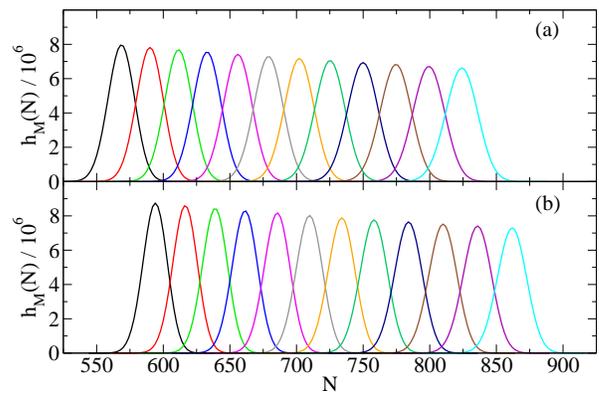}
\end{center}
\caption{(Color online) Some of the sampled length distributions for inverse temperature
 (a) $\beta=0$ (swollen phase) and (b) $\beta=0.5$ (collapsed phase).
The twelve curves are for different $M$ values, ranging from $M=690$ (left) to $M=1000$ (right).
The similar overlaps between contiguous distributions are obtained by choosing a
spacing between the corresponding $M$'s that  scales as $\sqrt{M}$.
}
\label{fig:hist}
\end{figure}

For multiple Markov chains at different $\beta$'s, 
the swapping of configurations between two contiguous (in $\beta$ space) 
Markov chain is likely to be accepted if there is a non negligible overlap 
between the energy distributions sampled by the two Markov chains.
Here the interacting Markov chains are taken at different $M$s and a good 
acceptance rate of swaps between two neighboring chains is expected if 
there is a non negligible overlap of the corresponding  $N$ distributions.
Since the dispersion in the $N$s sampled is $\sim \sqrt{M}$, the spacing
between subsequent values of $M$ should also scale as  $\sim \sqrt{M}$.
With this choice of $M$ the swapping moves should occur quite frequently,
allowing a good mobility of the sampled configuration in the space of $M$  values 
and hence decreasing the correlation times between in the stochastic sampling (See Fig.~\ref{fig:hist}).

\section{Results}

Simulations of circular IELP on the cubic lattice are based on the sampling scheme described
in the previous section with $K=1$, i.e.~$\omega = \mu$. 
By fixing $\beta$ we run in parallel 46 Markov chains,
each with a given value of $M$ from a minimum $M_{\min} = 100$ up to a maximum $M_{\max}=1000$.
As explained in section II, the asymptotic behavior of the average stored length density $\rho_M(\beta)$
can be exploited to investigate the scaling properties of the free energy of globular
rings. To test the validity of the approach, let us first consider the scaling of $\rho_M(\beta)$
in the swollen phase. In this case $\sigma = 0$ in~(\ref{ZN}) and the 
leading term in~(\ref{rho-fit}) is proportional to $\theta$ and  goes as $1/M$.

\begin{figure}[!tb]
\begin{center}
\includegraphics[width=7.8cm]{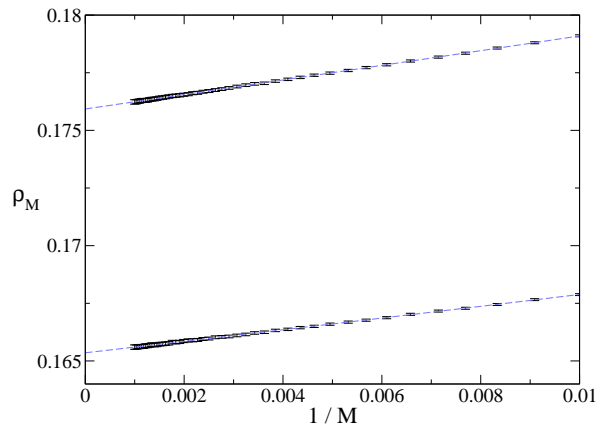}
\end{center}
\caption{(Color online) Mean stored length density vs $1/M$ 
for $\beta = 0$ (top) and $\beta = 0.269$ (bottom).
The dashed lines are fits based on (\ref{rho-fit}) with $\sigma=0$. 
}
\label{fig:rho1}
\end{figure}

Figure~\ref{fig:rho1} shows the scaling behavior of $\rho_M(0)$ as a function of $1/M$
for rings on the cubic lattice. The data converges linearly to an intercept $\rho_{\infty}$, as
expected. By fitting the data with $\Delta = 1/2$ we get the estimates
$\rho_{\infty}(0) \simeq 0.161253$ and $\theta \simeq -1.759$. Given that $\rho_{\infty}(\beta) = 1/[1+\mu(\beta)]$
and that in rings $\theta$ is usually denoted by $\alpha -2$, we finally get
\begin{equation}
\mu(0) = 4.68412(15),  \qquad  \alpha = 0.241 (22).
\end{equation}
These estimates are, within error bars, compatible with the ones obtained
in previous studies on the scaling properties of self-avoiding polygons~\cite{Madras&Slade:1993,Baiesi_et_al:2010:PRE}.
We perform a similar analysis at the $\Theta$-point 
($\beta=0.269$~\cite{Tesi_et_al:1996:J_Phys_A,Grassberger:1997:PRE}))
where the surface correction is still absent. The data are also reported in Figure~\ref{fig:rho1}
and by fitting them using~(\ref{rho-fit}) with $\Delta = 1/2$ and $\sigma=0$ we find
\begin{equation}
\mu(0.269) = 5.04762(22),  \qquad  \alpha = 0.528 (31).
\end{equation}
The estimate of $\alpha$ is more complicated than in the swollen phase due 
to the presence of logarithmic corrections~\cite{Duplantier:1982:JP} but the value we find agrees, 
within error bars, with the  mean field value $\alpha=1/2$ that, in $d=3$, is supposed to be exact.

\begin{figure}[!tb]
\begin{center}
\includegraphics[width=7.8cm]{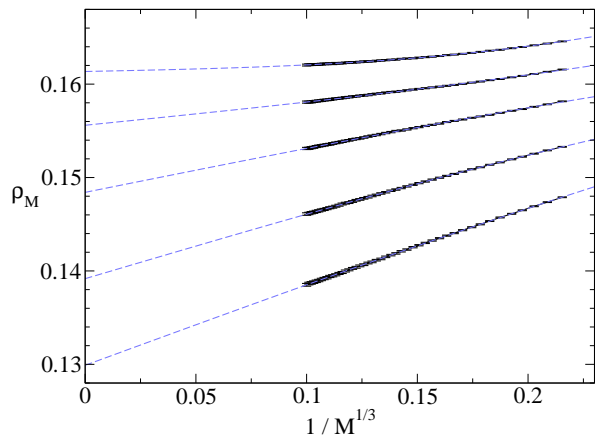}
\end{center}
\caption{(Color online) Mean stored length density vs $1/M^{1/3}$ 
for (from top to bottom) the values of $\beta = 0.32$, $0.36$, $0.4$, $0.45$, and $0.5$. 
The dashed lines are fits based on (\ref{rho-fit}) with $B=0$.
}
\label{fig:rho}
\end{figure}

We now focus on the scaling properties of globular rings by considering values of 
$\beta > \beta_{\Theta} = 0.269$~\cite{Grassberger:1997:PRE}.
In this case the surface free energy term proportional to $\sigma$ is not negligible 
and the  leading term in~(\ref{rho-fit}) should scale 
as $1/M^{1/3}$. This is confirmed in Figure~\ref{fig:rho} where,
by plotting the data $\rho_M(\beta)$ as a function of $1/M^{1/3}$ a linear behavior occurs 
for sufficiently large values of $M$ and $\beta$.
The linear extrapolation of these data at $1/M^{1/3}\to 0$ would 
give an estimate of $\rho_{\infty}(\beta)$, the slope of the linear approximation of the curves
an estimate of $\sigma(\beta)$ while their curvature  furnishes an estimate of $\theta$. 
Since finite-size effects are present in our data, one should in principle include into the 
non-linear fit the correction to scaling that depends on $B$ and $\Delta$. However, these 
two additional unknown parameters would destabilize the fit.
We decided instead to proceed as follows: 
we partitioned the data into three groups according to the $M$ values.
In the first group the data with the largest ten values of $M$ are excluded, in the second
group are excluded points with the five largest and the five smallest $M$s, while the third one
excluded the ten smallest $M$s. The third group, the most asymptotic one, has been used
to estimate $\mu(\beta)$, $\sigma(\beta)$, and $\theta$ while the estimates from the other
two groups are used to estimate the error bars.
In Table~\ref{tab:1} we  listed the estimates of $\mu(\beta)$, $\sigma(\beta)$ and
$\theta(\beta)$ for a wide values of $\beta$ in the globular phase. For comparison also the
estimates in the swollen phase and at the $\Theta$ point have been included.

\begin{table}[!b]
\begin{center}
\begin{tabular}{ l l l l l }
\hline
Phase &$\beta$ & $\mu$ & $\sigma$ & $\alpha$ \\
\hline
swollen            & 0 &        4.68412(15) &            & 0.241(22) \\
$\Theta$-point & 0.269 &        5.04762(22) &            & 0.528(31) \\
globular & 0.32 &       5.20(1) & 0.06(3) & 0.8(2) \\
         &0.36 &        5.45(4) &  0.34(10) & 2.4(9) \\
         &0.4  &        5.75(2) &  0.58(5) & 2.8(6) \\
         &0.45 &        6.17(2) &  0.82(4) & 2.5(3) \\
         &0.5  &        6.68(2) &  1.06(5) & 2.1(4) \\
\hline
\end{tabular}
\end{center}
\caption{Numerical estimates of $\mu$, $\sigma$ and $\alpha$ 
obtained by fitting the data of Figs.~\ref{fig:rho} and ~\ref{fig:rho1}
with Eq.~(\ref{rho-fit}).
}
\label{tab:1}
\end{table}

\begin{figure}[!tb]
\begin{center}
\includegraphics[width=7.4cm]{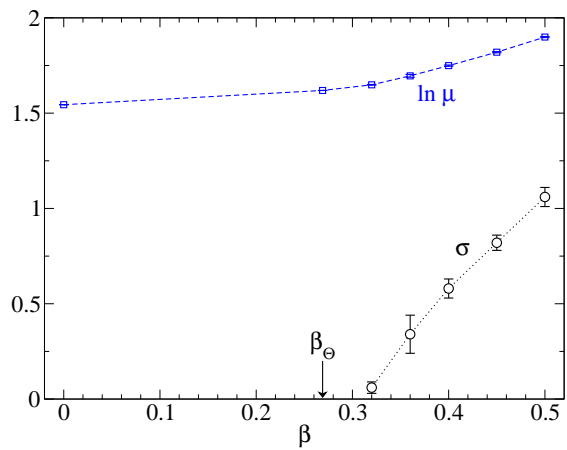}
\end{center}
\caption{(Color online) Estimate of the bulk ($\log \mu$)  and surface $\sigma$ free energy contributions  
at various inverse  temperatures  $\beta$.
}
\label{fig:sigma}
\end{figure}

We first notice that both $\log \mu(\beta)$ and $\sigma(\beta)$ in the globular phase
increase with $\beta$ (see Figure~\ref{fig:sigma})
and, deep inside the phase the dependence is linear in $\beta$. Since $\log \mu(\beta)$
is proportional to the bulk free energy per monomer $f(\beta)$ of the model, 
the linear behavior is expected by simple rigorous bounds on $f(\beta)$ 
(see for example~\cite{Tesi_et_al:1996:J_Phys_A}).
The  linear behavior of $\sigma(\beta)$ can be explained in a similar way
by noticing that, in the large $\beta$ limit, the statistics of the 
globular  phase is dominated by almost  ``spherical'' configurations  with 
surface area of the order of $C N^{2/3}$. 
This gives a bound on the surface free energy $\sigma(\beta)$ of the order of $\beta C$.

\begin{figure}[!tb]
\begin{center}
\includegraphics[width=7.8cm]{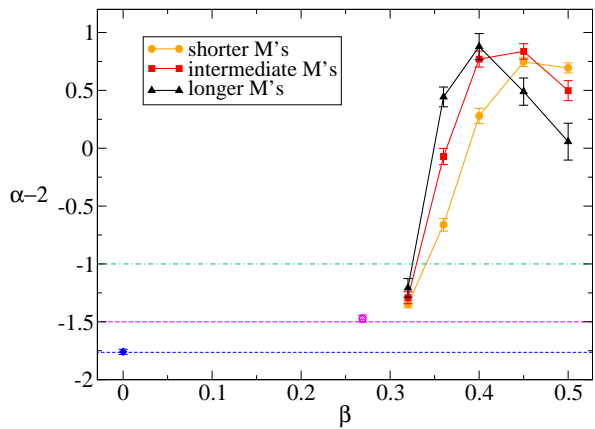}
\end{center}
\caption{(Color online) Estimates of the entropic exponent 
 $\alpha-2$ as a function of  the inverse temperatures $\beta$. 
The three sets of data for $\beta>0.3$ show how
the estimates of $\alpha$ change by taking more and more asymptotic data.
The dashed horizontal lines refer to the hyperscaling values $-\nu d$ 
expected respectively for  rings in the swollen phase and at the $\Theta$-point. 
The dot-dashed horizontal line refers  to the value that one would 
expect from hyperscaling in the
collapsed phase.
}
\label{fig:alpha}
\end{figure}

Finally the estimates of the entropic exponent $\alpha$ for rings are reported 
in the last column of Table~\ref{tab:1} and plotted in Fig.~\ref{fig:alpha} for the three sets of 
data grouped as explained before (we recall that the estimates at $\beta=0$ and $\beta=\beta_{\Theta}$ are 
obtained by using (\ref{rho-fit}) with $\sigma=0$ and $\Delta=1/2$). 
In addition, from the figure it is readily seen that the hyperscaling relation 
$2-\alpha = \nu d$~\cite{Vanderzande:1998,Duplantier_PRL86:branched} 
is confirmed both in the swollen phase ($\beta=0$), where $\nu \approx 0.58797(7)$~\cite{Clisby:2010:Phys-Rev-Lett} 
and at the $\Theta$ point, where $\nu= 1/2$. 
In the above relation $\nu$ is the critical exponent governing the scaling behavior of the 
average extension of the polymer,   and $d=3$.

In the globular phase ($\nu=1/d$) the value of $\alpha$ is not know from the theory and, 
as far as we know, there are no numerical results supporting the hyperscaling relation 
$\alpha-2 = - 1$.
The  plot in Figure~\ref{fig:alpha} shows  for $\beta >\beta_{\Theta}$ an initially steep increase 
of $\alpha-2$ followed by an unstable behavior deep in the globular phase. 
While the sharp increase just above the $\Theta$ point is a good indication of 
a single value of $\alpha$ characterizing the globular phase, the actual estimate of this value 
seems to fluctuate widely between  $\alpha=2$ and $\alpha = 2.7$. 
This is probably due either to the fact that the sampling at high values of $\beta$ 
is not robust and reliable  enough or that the neglected correction proportional to $B$ 
in (\ref{rho-fit}) is important in this region for the values of $M$ considered.
More investigations are needed to clarify this issue.
On the other hand, by focusing on the crossings formed by the three sets of 
estimates one  observe that, as $M$ 
increases, the $\beta$ values of the crossings shift towards the $\Theta$ point and correspondingly 
the $\alpha$ exponents goes closer to a value between $2.7$ and $3.0$.
Despite the important degree of uncertainty in the estimates of $\alpha$, we can 
say that, unlike in the swollen phase and at the $\Theta$ point, the hyperscaling relation
is not valid for globular polymers.

\section{Conclusions}
In this paper we have introduced and implemented a Monte Carlo scheme to study the 
scaling properties of $N$-steps interacting polymers on the cubic lattice.
The  sampling scheme is  based essentially on  the $N$-varying BFACF moves acting on a 
class of self-attracting rings that we refer as IELPs. 
These are chains of $M$ monomers that locally
could fold into the same lattice site and hence that
can accumulate their total length $M \ge N$ along their  $N$-steps backbone. 
If the interaction energy coincides with that of the backbone, however, it is not
necessary to simulate the moves of the $M$-monomer IELPs, it is just sufficient to sample $N$-step
configurations with a reweighting term that takes into account the multiplicity of the possible
rearrangements of $M$ monomers in $N$ sites.
Unlike the standard grand-canonical algorithms, 
this scheme bounds from above the number of steps $N$ by $M$, preventing the uncontrolled 
growth of the average $\mean{N}$ during the sampling, a problem that severely  affects 
the $N$-varying sampling of globular configurations.

By looking at the asymptotic properties of the stored length density $\rho_M(\beta)$
as a function of the inverse temperature $\beta$ we have been able to study the scaling 
behavior of the free energy of the 
interacting SAP model for several values of $\beta$,  from the swollen phase 
down to the globular phase, passing through the $\Theta$ point. 
To test the validity of our algorithm, we first estimated the bulk free energy $\log\mu(\beta)$ and the entropic 
exponent $\alpha-2$ of the model at $\beta=0$ and $\beta=\beta_{\Theta}$, finding a good agreement with 
previous results.
In the so far unexplored globular phase, in addition to the bulk free energy $\log(\mu)$ 
we gave an estimate of the surface free energy $\sigma(\beta)$.
Finally, by looking at the entropic exponent $\alpha-2$ we furnish good evidence that, unlike in the swollen 
phase and at the $\Theta$ point, the hyperscaling relation $\alpha -2 = -\nu d$ 
does not hold  for the class of collapsed globular rings.

The Monte Carlo scheme based on IELPs
should be particularly  useful to study globular rings of fixed knot type~\cite{Baiesi_et_al:2010:PRE}.
It would then be  interesting to apply this sampling scheme to investigate the role of topology in the 
scaling properties of the free energy of globular knotted rings.

\acknowledgements
We thank E. Carlon and A. Stella for useful discussions.
We acknowledge financial support from the Italian Ministry of Education (Grant PRIN 2010HXAW77).


\begin{thebibliography}{38}%
\makeatletter
\providecommand \@ifxundefined [1]{%
 \@ifx{#1\undefined}
}%
\providecommand \@ifnum [1]{%
 \ifnum #1\expandafter \@firstoftwo
 \else \expandafter \@secondoftwo
 \fi
}%
\providecommand \@ifx [1]{%
 \ifx #1\expandafter \@firstoftwo
 \else \expandafter \@secondoftwo
 \fi
}%
\providecommand \natexlab [1]{#1}%
\providecommand \enquote  [1]{``#1''}%
\providecommand \bibnamefont  [1]{#1}%
\providecommand \bibfnamefont [1]{#1}%
\providecommand \citenamefont [1]{#1}%
\providecommand \href@noop [0]{\@secondoftwo}%
\providecommand \href [0]{\begingroup \@sanitize@url \@href}%
\providecommand \@href[1]{\@@startlink{#1}\@@href}%
\providecommand \@@href[1]{\endgroup#1\@@endlink}%
\providecommand \@sanitize@url [0]{\catcode `\\12\catcode `\$12\catcode
  `\&12\catcode `\#12\catcode `\^12\catcode `\_12\catcode `\%12\relax}%
\providecommand \@@startlink[1]{}%
\providecommand \@@endlink[0]{}%
\providecommand \url  [0]{\begingroup\@sanitize@url \@url }%
\providecommand \@url [1]{\endgroup\@href {#1}{\urlprefix }}%
\providecommand \urlprefix  [0]{URL }%
\providecommand \Eprint [0]{\href }%
\providecommand \doibase [0]{http://dx.doi.org/}%
\providecommand \selectlanguage [0]{\@gobble}%
\providecommand \bibinfo  [0]{\@secondoftwo}%
\providecommand \bibfield  [0]{\@secondoftwo}%
\providecommand \translation [1]{[#1]}%
\providecommand \BibitemOpen [0]{}%
\providecommand \bibitemStop [0]{}%
\providecommand \bibitemNoStop [0]{.\EOS\space}%
\providecommand \EOS [0]{\spacefactor3000\relax}%
\providecommand \BibitemShut  [1]{\csname bibitem#1\endcsname}%
\let\auto@bib@innerbib\@empty
\bibitem [{\citenamefont {{de Gennes}}(1979)}]{DeGennes:1979}%
  \BibitemOpen
  \bibfield  {author} {\bibinfo {author} {\bibfnamefont {P.-G.}\ \bibnamefont
  {{de Gennes}}},\ }\href@noop {} {\emph {\bibinfo {title} {Scaling concepts in
  Polymer Physics}}}\ (\bibinfo  {publisher} {Cornell University Press, Ithaca,
  New York},\ \bibinfo {year} {1979})\BibitemShut {NoStop}%
\bibitem [{\citenamefont {Vanderzande}(1998)}]{Vanderzande:1998}%
  \BibitemOpen
  \bibfield  {author} {\bibinfo {author} {\bibfnamefont {C.}~\bibnamefont
  {Vanderzande}},\ }\href@noop {} {\emph {\bibinfo {title} {Lattice models of
  Polymers}}}\ (\bibinfo  {publisher} {Cambridge University Press},\ \bibinfo
  {address} {Cambridge},\ \bibinfo {year} {1998})\BibitemShut {NoStop}%
\bibitem [{\citenamefont {{Janse van Rensburg}}(2000)}]{JansevanRensburg_book}%
  \BibitemOpen
  \bibfield  {author} {\bibinfo {author} {\bibfnamefont {E.~J.}\ \bibnamefont
  {{Janse van Rensburg}}},\ }\href@noop {} {\emph {\bibinfo {title}
  {Statistical Mechanics of Interacting Walks, Polygons, Animals and
  Vesicles}}}\ (\bibinfo  {publisher} {Oxford University Press},\ \bibinfo
  {address} {Oxford},\ \bibinfo {year} {2000})\BibitemShut {NoStop}%
\bibitem [{\citenamefont {Duplantier}\ and\ \citenamefont
  {Saleur}(1987{\natexlab{a}})}]{DuplantierSaleur_PRL87:Theta2d}%
  \BibitemOpen
  \bibfield  {author} {\bibinfo {author} {\bibfnamefont {B.}~\bibnamefont
  {Duplantier}}\ and\ \bibinfo {author} {\bibfnamefont {H.}~\bibnamefont
  {Saleur}},\ }\href@noop {} {\bibfield  {journal} {\bibinfo  {journal} {Phys.\
  Rev.\ Lett.}\ }\textbf {\bibinfo {volume} {59}},\ \bibinfo {pages} {539}
  (\bibinfo {year} {1987}{\natexlab{a}})}\BibitemShut {NoStop}%
\bibitem [{\citenamefont {Seno}\ and\ \citenamefont
  {Stella}(1988)}]{Seno_Stella:1988}%
  \BibitemOpen
  \bibfield  {author} {\bibinfo {author} {\bibfnamefont {F.}~\bibnamefont
  {Seno}}\ and\ \bibinfo {author} {\bibfnamefont {A.~L.}\ \bibnamefont
  {Stella}},\ }\href@noop {} {\bibfield  {journal} {\bibinfo  {journal} {J.
  Physique}\ }\textbf {\bibinfo {volume} {49}},\ \bibinfo {pages} {739}
  (\bibinfo {year} {1988})}\BibitemShut {NoStop}%
\bibitem [{\citenamefont {Grassberger}\ and\ \citenamefont
  {Hegger}(1995)}]{GrassbergerHegger_JP95:theta2d}%
  \BibitemOpen
  \bibfield  {author} {\bibinfo {author} {\bibfnamefont {P.}~\bibnamefont
  {Grassberger}}\ and\ \bibinfo {author} {\bibfnamefont {R.}~\bibnamefont
  {Hegger}},\ }\href@noop {} {\bibfield  {journal} {\bibinfo  {journal} {J.
  Phys. I France}\ }\textbf {\bibinfo {volume} {5}},\ \bibinfo {pages} {597}
  (\bibinfo {year} {1995})}\BibitemShut {NoStop}%
\bibitem [{\citenamefont {Tesi}\ \emph
  {et~al.}(1996{\natexlab{a}})\citenamefont {Tesi}, \citenamefont {{Janse van
  Rensburg}}, \citenamefont {Orlandini},\ and\ \citenamefont
  {Whittington}}]{Tesi_et_al:1996:J_Stat_Phys}%
  \BibitemOpen
  \bibfield  {author} {\bibinfo {author} {\bibfnamefont {M.}~\bibnamefont
  {Tesi}}, \bibinfo {author} {\bibfnamefont {E.~J.}\ \bibnamefont {{Janse van
  Rensburg}}}, \bibinfo {author} {\bibfnamefont {E.}~\bibnamefont {Orlandini}},
  \ and\ \bibinfo {author} {\bibfnamefont {S.~G.}\ \bibnamefont
  {Whittington}},\ }\href@noop {} {\bibfield  {journal} {\bibinfo  {journal}
  {J. Stat. Phys.}\ }\textbf {\bibinfo {volume} {82}},\ \bibinfo {pages} {155}
  (\bibinfo {year} {1996}{\natexlab{a}})}\BibitemShut {NoStop}%
\bibitem [{\citenamefont {Iwata}(1989)}]{Iwata:1989:Macromol}%
  \BibitemOpen
  \bibfield  {author} {\bibinfo {author} {\bibfnamefont {K.}~\bibnamefont
  {Iwata}},\ }\href@noop {} {\bibfield  {journal} {\bibinfo  {journal}
  {Macromol.}\ }\textbf {\bibinfo {volume} {22}},\ \bibinfo {pages} {3702}
  (\bibinfo {year} {1989})}\BibitemShut {NoStop}%
\bibitem [{\citenamefont {Owczarek}\ \emph
  {et~al.}(1993{\natexlab{a}})\citenamefont {Owczarek}, \citenamefont
  {Prellberg},\ and\ \citenamefont {Brak}}]{Owczarek_PB_PRL93:collapsed}%
  \BibitemOpen
  \bibfield  {author} {\bibinfo {author} {\bibfnamefont {A.~L.}\ \bibnamefont
  {Owczarek}}, \bibinfo {author} {\bibfnamefont {T.}~\bibnamefont {Prellberg}},
  \ and\ \bibinfo {author} {\bibfnamefont {R.}~\bibnamefont {Brak}},\
  }\href@noop {} {\bibfield  {journal} {\bibinfo  {journal} {Phys.\ Rev.\
  Lett.}\ }\textbf {\bibinfo {volume} {70}},\ \bibinfo {pages} {951} (\bibinfo
  {year} {1993}{\natexlab{a}})}\BibitemShut {NoStop}%
\bibitem [{\citenamefont {Duplantier}(1993)}]{Duplantier_PRL93:comment}%
  \BibitemOpen
  \bibfield  {author} {\bibinfo {author} {\bibfnamefont {B.}~\bibnamefont
  {Duplantier}},\ }\href@noop {} {\bibfield  {journal} {\bibinfo  {journal}
  {Phys.\ Rev.\ Lett.}\ }\textbf {\bibinfo {volume} {71}},\ \bibinfo {pages}
  {4274} (\bibinfo {year} {1993})}\BibitemShut {NoStop}%
\bibitem [{\citenamefont {Owczarek}\ \emph
  {et~al.}(1993{\natexlab{b}})\citenamefont {Owczarek}, \citenamefont
  {Prellberg},\ and\ \citenamefont {Brak}}]{Owczarek_PB_PRL93:reply}%
  \BibitemOpen
  \bibfield  {author} {\bibinfo {author} {\bibfnamefont {A.~L.}\ \bibnamefont
  {Owczarek}}, \bibinfo {author} {\bibfnamefont {T.}~\bibnamefont {Prellberg}},
  \ and\ \bibinfo {author} {\bibfnamefont {R.}~\bibnamefont {Brak}},\
  }\href@noop {} {\bibfield  {journal} {\bibinfo  {journal} {Phys.\ Rev.\
  Lett.}\ }\textbf {\bibinfo {volume} {71}},\ \bibinfo {pages} {4275} (\bibinfo
  {year} {1993}{\natexlab{b}})}\BibitemShut {NoStop}%
\bibitem [{\citenamefont {Brak}\ \emph {et~al.}(1993)\citenamefont {Brak},
  \citenamefont {Owczarek},\ and\ \citenamefont {Prellberg}}]{Brak:1993:JPA}%
  \BibitemOpen
  \bibfield  {author} {\bibinfo {author} {\bibfnamefont {R.}~\bibnamefont
  {Brak}}, \bibinfo {author} {\bibfnamefont {A.~L.}\ \bibnamefont {Owczarek}},
  \ and\ \bibinfo {author} {\bibfnamefont {T.}~\bibnamefont {Prellberg}},\
  }\href@noop {} {\bibfield  {journal} {\bibinfo  {journal} {J. Phys. A.}\
  }\textbf {\bibinfo {volume} {26}},\ \bibinfo {pages} {4565} (\bibinfo {year}
  {1993})}\BibitemShut {NoStop}%
\bibitem [{\citenamefont {Grassberger}\ and\ \citenamefont
  {Hsu}(2002)}]{Grassberger:PRE:2002}%
  \BibitemOpen
  \bibfield  {author} {\bibinfo {author} {\bibfnamefont {P.}~\bibnamefont
  {Grassberger}}\ and\ \bibinfo {author} {\bibfnamefont {H.~P.}\ \bibnamefont
  {Hsu}},\ }\href@noop {} {\bibfield  {journal} {\bibinfo  {journal} {Phys.\
  Rev.\ E}\ }\textbf {\bibinfo {volume} {65}},\ \bibinfo {pages} {031807}
  (\bibinfo {year} {2002})}\BibitemShut {NoStop}%
\bibitem [{\citenamefont {Baiesi}\ \emph {et~al.}(2006)\citenamefont {Baiesi},
  \citenamefont {Orlandini},\ and\ \citenamefont {Stella}}]{Baiesi:2006:PRL}%
  \BibitemOpen
  \bibfield  {author} {\bibinfo {author} {\bibfnamefont {M.}~\bibnamefont
  {Baiesi}}, \bibinfo {author} {\bibfnamefont {E.}~\bibnamefont {Orlandini}}, \
  and\ \bibinfo {author} {\bibfnamefont {A.~L.}\ \bibnamefont {Stella}},\
  }\href@noop {} {\bibfield  {journal} {\bibinfo  {journal} {Phys.\ Rev.\
  Lett.}\ }\textbf {\bibinfo {volume} {96}},\ \bibinfo {pages} {040602}
  (\bibinfo {year} {2006})}\BibitemShut {NoStop}%
\bibitem [{\citenamefont {Baiesi}\ \emph {et~al.}(2007)\citenamefont {Baiesi},
  \citenamefont {Orlandini},\ and\ \citenamefont {Stella}}]{Baiesi:2007:PRL}%
  \BibitemOpen
  \bibfield  {author} {\bibinfo {author} {\bibfnamefont {M.}~\bibnamefont
  {Baiesi}}, \bibinfo {author} {\bibfnamefont {E.}~\bibnamefont {Orlandini}}, \
  and\ \bibinfo {author} {\bibfnamefont {A.~L.}\ \bibnamefont {Stella}},\
  }\href@noop {} {\bibfield  {journal} {\bibinfo  {journal} {Phys.\ Rev.\
  Lett.}\ }\textbf {\bibinfo {volume} {99}},\ \bibinfo {pages} {058301}
  (\bibinfo {year} {2007})}\BibitemShut {NoStop}%
\bibitem [{\citenamefont {Brak}\ \emph {et~al.}(1992)\citenamefont {Brak},
  \citenamefont {Guttmann},\ and\ \citenamefont {Whittington}}]{Brak:1992:JPA}%
  \BibitemOpen
  \bibfield  {author} {\bibinfo {author} {\bibfnamefont {R.}~\bibnamefont
  {Brak}}, \bibinfo {author} {\bibfnamefont {A.~J.}\ \bibnamefont {Guttmann}},
  \ and\ \bibinfo {author} {\bibfnamefont {S.}~\bibnamefont {Whittington}},\
  }\href@noop {} {\bibfield  {journal} {\bibinfo  {journal} {J.\ Phys. A: Math.
  Gen.}\ }\textbf {\bibinfo {volume} {25}},\ \bibinfo {pages} {2437} (\bibinfo
  {year} {1992})}\BibitemShut {NoStop}%
\bibitem [{\citenamefont {Nguyen}\ and\ \citenamefont
  {P\'etr\'elis}(2013)}]{Nguyen:2013:JSP}%
  \BibitemOpen
  \bibfield  {author} {\bibinfo {author} {\bibfnamefont {G.~B.}\ \bibnamefont
  {Nguyen}}\ and\ \bibinfo {author} {\bibfnamefont {N.}~\bibnamefont
  {P\'etr\'elis}},\ }\href@noop {} {\bibfield  {journal} {\bibinfo  {journal}
  {J. Stat. Phys.}\ }\textbf {\bibinfo {volume} {151}},\ \bibinfo {pages}
  {1099} (\bibinfo {year} {2013})}\BibitemShut {NoStop}%
\bibitem [{\citenamefont {Duplantier}\ and\ \citenamefont
  {Saleur}(1987{\natexlab{b}})}]{DuplantierSaleur_NPB87:dense}%
  \BibitemOpen
  \bibfield  {author} {\bibinfo {author} {\bibfnamefont {B.}~\bibnamefont
  {Duplantier}}\ and\ \bibinfo {author} {\bibfnamefont {H.}~\bibnamefont
  {Saleur}},\ }\href@noop {} {\bibfield  {journal} {\bibinfo  {journal} {Nucl.\
  Phys.\ B}\ }\textbf {\bibinfo {volume} {290 [FS20]}},\ \bibinfo {pages} {291}
  (\bibinfo {year} {1987}{\natexlab{b}})}\BibitemShut {NoStop}%
\bibitem [{\citenamefont {Duplantier}\ and\ \citenamefont
  {David}(1988)}]{DuplantierDavid_JSP88}%
  \BibitemOpen
  \bibfield  {author} {\bibinfo {author} {\bibfnamefont {B.}~\bibnamefont
  {Duplantier}}\ and\ \bibinfo {author} {\bibfnamefont {F.}~\bibnamefont
  {David}},\ }\href@noop {} {\bibfield  {journal} {\bibinfo  {journal} {J.\
  Stat.\ Phys.}\ }\textbf {\bibinfo {volume} {51}},\ \bibinfo {pages} {327}
  (\bibinfo {year} {1988})}\BibitemShut {NoStop}%
\bibitem [{\citenamefont {Berg}\ and\ \citenamefont
  {Foester}(1981)}]{Berg&Foester:1981:Phys-Lett-B}%
  \BibitemOpen
  \bibfield  {author} {\bibinfo {author} {\bibfnamefont {B.}~\bibnamefont
  {Berg}}\ and\ \bibinfo {author} {\bibfnamefont {D.}~\bibnamefont {Foester}},\
  }\href@noop {} {\bibfield  {journal} {\bibinfo  {journal} {Phys. Lett. B}\
  }\textbf {\bibinfo {volume} {106}},\ \bibinfo {pages} {323} (\bibinfo {year}
  {1981})}\BibitemShut {NoStop}%
\bibitem [{\citenamefont {de~Carvalho}\ and\ \citenamefont
  {Caracciolo}(1983)}]{Aragao&Caracciolo:1983:J-Physique}%
  \BibitemOpen
  \bibfield  {author} {\bibinfo {author} {\bibfnamefont {C.~A.}\ \bibnamefont
  {de~Carvalho}}\ and\ \bibinfo {author} {\bibfnamefont {S.}~\bibnamefont
  {Caracciolo}},\ }\href@noop {} {\bibfield  {journal} {\bibinfo  {journal} {J.
  Physique}\ }\textbf {\bibinfo {volume} {44}},\ \bibinfo {pages} {323}
  (\bibinfo {year} {1983})}\BibitemShut {NoStop}%
\bibitem [{\citenamefont {Grassberger}(1997)}]{Grassberger:1997:PRE}%
  \BibitemOpen
  \bibfield  {author} {\bibinfo {author} {\bibfnamefont {P.}~\bibnamefont
  {Grassberger}},\ }\href@noop {} {\bibfield  {journal} {\bibinfo  {journal}
  {Phys. Rev. E}\ }\textbf {\bibinfo {volume} {56}},\ \bibinfo {pages} {3682}
  (\bibinfo {year} {1997})}\BibitemShut {NoStop}%
\bibitem [{\citenamefont {Rechnitzer}\ and\ \citenamefont {{Janse van
  Rensburg}}(2008)}]{Rechnitzer:2008:JPA}%
  \BibitemOpen
  \bibfield  {author} {\bibinfo {author} {\bibfnamefont {A.}~\bibnamefont
  {Rechnitzer}}\ and\ \bibinfo {author} {\bibfnamefont {E.~J.}\ \bibnamefont
  {{Janse van Rensburg}}},\ }\href@noop {} {\bibfield  {journal} {\bibinfo
  {journal} {J.\ Phys. A: Math. Gen.}\ }\textbf {\bibinfo {volume} {41}},\
  \bibinfo {pages} {442002} (\bibinfo {year} {2008})}\BibitemShut {NoStop}%
\bibitem [{\citenamefont {{Janse van Rensburg}}\ and\ \citenamefont
  {Whittington}(1991)}]{Janse-van-Rensburg&Whittington:1991b:J-Phys-A}%
  \BibitemOpen
  \bibfield  {author} {\bibinfo {author} {\bibfnamefont {E.~J.}\ \bibnamefont
  {{Janse van Rensburg}}}\ and\ \bibinfo {author} {\bibfnamefont {S.~G.}\
  \bibnamefont {Whittington}},\ }\href@noop {} {\bibfield  {journal} {\bibinfo
  {journal} {J. Phys. A: Math. Gen.}\ }\textbf {\bibinfo {volume} {24}},\
  \bibinfo {pages} {5553} (\bibinfo {year} {1991})}\BibitemShut {NoStop}%
\bibitem [{\citenamefont {Orlandini}\ \emph {et~al.}(1998)\citenamefont
  {Orlandini}, \citenamefont {Tesi}, \citenamefont {{Janse van Rensburg}},\
  and\ \citenamefont {Whittington}}]{Orlandini:1998:J-Phys-A}%
  \BibitemOpen
  \bibfield  {author} {\bibinfo {author} {\bibfnamefont {E.}~\bibnamefont
  {Orlandini}}, \bibinfo {author} {\bibfnamefont {M.}~\bibnamefont {Tesi}},
  \bibinfo {author} {\bibfnamefont {E.~J.}\ \bibnamefont {{Janse van
  Rensburg}}}, \ and\ \bibinfo {author} {\bibfnamefont {S.~G.}\ \bibnamefont
  {Whittington}},\ }\href@noop {} {\bibfield  {journal} {\bibinfo  {journal}
  {J. Phys. A: Math. Gen.}\ }\textbf {\bibinfo {volume} {31}},\ \bibinfo
  {pages} {5953} (\bibinfo {year} {1998})}\BibitemShut {NoStop}%
\bibitem [{\citenamefont {Marcone}\ \emph {et~al.}(2007)\citenamefont
  {Marcone}, \citenamefont {Orlandini}, \citenamefont {Stella},\ and\
  \citenamefont {Zonta}}]{Marcone:2007:PRE}%
  \BibitemOpen
  \bibfield  {author} {\bibinfo {author} {\bibfnamefont {B.}~\bibnamefont
  {Marcone}}, \bibinfo {author} {\bibfnamefont {E.}~\bibnamefont {Orlandini}},
  \bibinfo {author} {\bibfnamefont {A.~L.}\ \bibnamefont {Stella}}, \ and\
  \bibinfo {author} {\bibfnamefont {F.}~\bibnamefont {Zonta}},\ }\href@noop {}
  {\bibfield  {journal} {\bibinfo  {journal} {Phys.\ Rev.\ E}\ }\textbf
  {\bibinfo {volume} {75}},\ \bibinfo {pages} {041105} (\bibinfo {year}
  {2007})}\BibitemShut {NoStop}%
\bibitem [{\citenamefont {Baiesi}\ \emph {et~al.}(2011)\citenamefont {Baiesi},
  \citenamefont {Orlandini}, \citenamefont {Stella},\ and\ \citenamefont
  {Zonta}}]{Baiesi_et_al:2011:PRL}%
  \BibitemOpen
  \bibfield  {author} {\bibinfo {author} {\bibfnamefont {M.}~\bibnamefont
  {Baiesi}}, \bibinfo {author} {\bibfnamefont {E.}~\bibnamefont {Orlandini}},
  \bibinfo {author} {\bibfnamefont {A.~L.}\ \bibnamefont {Stella}}, \ and\
  \bibinfo {author} {\bibfnamefont {F.}~\bibnamefont {Zonta}},\ }\href@noop {}
  {\bibfield  {journal} {\bibinfo  {journal} {Phys.\ Rev.\ Lett.}\ }\textbf
  {\bibinfo {volume} {106}},\ \bibinfo {pages} {258301} (\bibinfo {year}
  {2011})}\BibitemShut {NoStop}%
\bibitem [{\citenamefont {van Heukelum}\ and\ \citenamefont
  {Barkema}(2003)}]{Heukelum:2003:JChemPhys}%
  \BibitemOpen
  \bibfield  {author} {\bibinfo {author} {\bibfnamefont {A.}~\bibnamefont {van
  Heukelum}}\ and\ \bibinfo {author} {\bibfnamefont {G.~T.}\ \bibnamefont
  {Barkema}},\ }\href@noop {} {\bibfield  {journal} {\bibinfo  {journal} {J.
  Chem. Phys.}\ }\textbf {\bibinfo {volume} {119}},\ \bibinfo {pages} {8197}
  (\bibinfo {year} {2003})}\BibitemShut {NoStop}%
\bibitem [{\citenamefont {Wolterink}\ \emph {et~al.}(2006)\citenamefont
  {Wolterink}, \citenamefont {Barkema},\ and\ \citenamefont
  {Panja}}]{Wolterink:2006:PRL}%
  \BibitemOpen
  \bibfield  {author} {\bibinfo {author} {\bibfnamefont {J.~K.}\ \bibnamefont
  {Wolterink}}, \bibinfo {author} {\bibfnamefont {G.~T.}\ \bibnamefont
  {Barkema}}, \ and\ \bibinfo {author} {\bibfnamefont {D.}~\bibnamefont
  {Panja}},\ }\href@noop {} {\bibfield  {journal} {\bibinfo  {journal} {Phys.
  Rev. Lett.}\ }\textbf {\bibinfo {volume} {96}},\ \bibinfo {pages} {208301}
  (\bibinfo {year} {2006})}\BibitemShut {NoStop}%
\bibitem [{\citenamefont {Baiesi}\ \emph {et~al.}(2010)\citenamefont {Baiesi},
  \citenamefont {Barkema},\ and\ \citenamefont
  {Carlon}}]{Baiesi_et_al:2010:PRE}%
  \BibitemOpen
  \bibfield  {author} {\bibinfo {author} {\bibfnamefont {M.}~\bibnamefont
  {Baiesi}}, \bibinfo {author} {\bibfnamefont {G.~T.}\ \bibnamefont {Barkema}},
  \ and\ \bibinfo {author} {\bibfnamefont {E.}~\bibnamefont {Carlon}},\
  }\href@noop {} {\bibfield  {journal} {\bibinfo  {journal} {Phys.\ Rev.\ E}\
  }\textbf {\bibinfo {volume} {81}},\ \bibinfo {pages} {061801} (\bibinfo
  {year} {2010})}\BibitemShut {NoStop}%
\bibitem [{\citenamefont {Schram}\ \emph {et~al.}(2013)\citenamefont {Schram},
  \citenamefont {Barkema},\ and\ \citenamefont {Schiessel}}]{Schram:JCP:2013}%
  \BibitemOpen
  \bibfield  {author} {\bibinfo {author} {\bibfnamefont {R.~D.}\ \bibnamefont
  {Schram}}, \bibinfo {author} {\bibfnamefont {G.~T.}\ \bibnamefont {Barkema}},
  \ and\ \bibinfo {author} {\bibfnamefont {H.}~\bibnamefont {Schiessel}},\
  }\href@noop {} {\bibfield  {journal} {\bibinfo  {journal} {J. Chem. Phys.}\
  }\textbf {\bibinfo {volume} {138}},\ \bibinfo {pages} {224901} (\bibinfo
  {year} {2013})}\BibitemShut {NoStop}%
\bibitem [{\citenamefont {{N. Madras and G. Slade}}(1993)}]{Madras&Slade:1993}%
  \BibitemOpen
  \bibfield  {author} {\bibinfo {author} {\bibnamefont {{N. Madras and G.
  Slade}}},\ }\href@noop {} {\emph {\bibinfo {title} {{The {Self}-{Avoiding}
  {Walk}}}}}\ (\bibinfo  {publisher} {{Birkh\"auser}},\ \bibinfo {address}
  {Berlin},\ \bibinfo {year} {{1993}})\BibitemShut {NoStop}%
\bibitem [{\citenamefont {Meyer}(1991)}]{Meyer:1991}%
  \BibitemOpen
  \bibfield  {author} {\bibinfo {author} {\bibfnamefont {C.~J.}\ \bibnamefont
  {Meyer}},\ }in\ \href@noop {} {\emph {\bibinfo {booktitle} {Computing Science
  and Statistics: Proceedings of the 23rd Symposium on the Interface}}},\
  \bibinfo {editor} {edited by\ \bibinfo {editor} {\bibfnamefont {E.~M.}\
  \bibnamefont {Keramides}}}\ (\bibinfo  {publisher} {Interface Foundation},\
  \bibinfo {address} {Fairfax Station, Virginia},\ \bibinfo {year} {1991})\
  pp.\ \bibinfo {pages} {156--163}\BibitemShut {NoStop}%
\bibitem [{\citenamefont {Orlandini}(1998)}]{Orlandini:1998:IMA}%
  \BibitemOpen
  \bibfield  {author} {\bibinfo {author} {\bibfnamefont {E.}~\bibnamefont
  {Orlandini}},\ }\href@noop {} {\bibfield  {journal} {\bibinfo  {journal}
  {Numerical Methods for Polymeric Systems, edited by S. G. Whittington, IMA
  Volumes in Mathematics and Its Application}\ }\textbf {\bibinfo {volume}
  {102}},\ \bibinfo {pages} {33} (\bibinfo {year} {1998})}\BibitemShut
  {NoStop}%
\bibitem [{\citenamefont {Tesi}\ \emph
  {et~al.}(1996{\natexlab{b}})\citenamefont {Tesi}, \citenamefont {{Janse van
  Rensburg}}, \citenamefont {Orlandini},\ and\ \citenamefont
  {Whittington}}]{Tesi_et_al:1996:J_Phys_A}%
  \BibitemOpen
  \bibfield  {author} {\bibinfo {author} {\bibfnamefont {M.}~\bibnamefont
  {Tesi}}, \bibinfo {author} {\bibfnamefont {E.~J.}\ \bibnamefont {{Janse van
  Rensburg}}}, \bibinfo {author} {\bibfnamefont {E.}~\bibnamefont {Orlandini}},
  \ and\ \bibinfo {author} {\bibfnamefont {S.~G.}\ \bibnamefont
  {Whittington}},\ }\href@noop {} {\bibfield  {journal} {\bibinfo  {journal}
  {J.\ Phys. A: Math. Gen.}\ }\textbf {\bibinfo {volume} {29}},\ \bibinfo
  {pages} {2451} (\bibinfo {year} {1996}{\natexlab{b}})}\BibitemShut {NoStop}%
\bibitem [{\citenamefont {Duplantier}(1982)}]{Duplantier:1982:JP}%
  \BibitemOpen
  \bibfield  {author} {\bibinfo {author} {\bibfnamefont {B.}~\bibnamefont
  {Duplantier}},\ }\href@noop {} {\bibfield  {journal} {\bibinfo  {journal} {J.
  Physique}\ }\textbf {\bibinfo {volume} {43}},\ \bibinfo {pages} {991}
  (\bibinfo {year} {1982})}\BibitemShut {NoStop}%
\bibitem [{\citenamefont {Duplantier}(1986)}]{Duplantier_PRL86:branched}%
  \BibitemOpen
  \bibfield  {author} {\bibinfo {author} {\bibfnamefont {B.}~\bibnamefont
  {Duplantier}},\ }\href@noop {} {\bibfield  {journal} {\bibinfo  {journal}
  {Phys.\ Rev.\ Lett.}\ }\textbf {\bibinfo {volume} {57}},\ \bibinfo {pages}
  {941} (\bibinfo {year} {1986})}\BibitemShut {NoStop}%
\bibitem [{\citenamefont {Clisby}(2010)}]{Clisby:2010:Phys-Rev-Lett}%
  \BibitemOpen
  \bibfield  {author} {\bibinfo {author} {\bibfnamefont {N.}~\bibnamefont
  {Clisby}},\ }\href@noop {} {\bibfield  {journal} {\bibinfo  {journal} {Phys.
  Rev. Lett.}\ }\textbf {\bibinfo {volume} {104}},\ \bibinfo {pages} {055702}
  (\bibinfo {year} {2010})}\BibitemShut {NoStop}%
\end{thebibliography}

%

\end{document}